\documentclass[11pt,aps,preprint,showpacs,amsmath,amssymb,amsfonts,eqnarray,array]{revtex4}
\newcommand{\overbar}[1]{\mkern 1.5mu\overline{\mkern-2.5mu#1\mkern-1.5mu}\mkern 1.5mu}
\usepackage{anysize}
\marginsize{1.9cm}{1.9cm}{1.55cm}{1.55cm}
\usepackage{graphicx}
\usepackage{dcolumn}
\usepackage{bm}
\begin{document}


\title{Magneto-optical absorption in semiconducting spherical quantum dots: Influence of the dot-size,
confining potential, and magnetic field \\}
\author{Manvir S. Kushwaha}
\affiliation
{Department of Physics and Astronomy, Rice University, P.O. Box 1892, Houston, TX 77251, USA}
\date{\today}

\begin{abstract}

Semiconducting quantum dots -- more fancifully dubbed artificial atoms -- are quasi-zero dimensional,
tiny, man-made systems with charge carriers {\em completely} confined in all three dimensions. The
scientific quest behind the synthesis of quantum dots is to create and control future electronic and
optical nanostructures engineered through tailoring size, shape, and composition.
The {\em complete} confinement -- or the lack of any degree of freedom for the electrons (and/or holes)
-- in quantum dots limits the exploration of {\em spatially localized} elementary excitations such as
plasmons to {\em direct} rather than {\em reciprocal} space.
Here we embark on a thorough investigation of
the magneto-optical absorption in semiconducting {\em spherical} quantum dots
characterized by a confining harmonic potential and an applied magnetic field in the symmetric gauge.
This is done within the framework of Bohm-Pines' random-phase approximation that enables us to derive
and discuss the full Dyson equation that takes proper account of the Coulomb interactions.
As an application of our theoretical strategy, we compute various single-particle and many-particle
phenomena such as the Fock-Darwin spectrum; Fermi energy; magneto-optical transitions; probability
distribution; and the magneto-optical absorption in the quantum dots.
It is observed that the role of an applied magnetic field on the absorption spectrum is comparable
to that of a confining potential. Increasing (decreasing) the strength of the magnetic field or the
confining potential is found to be analogous to shrinking (expanding) the size of the quantum dots:
resulting into a blue (red) shift in the absorption spectrum. The Fermi energy diminishes with both
increasing magnetic-field and dot-size; and exhibits saw-tooth-like oscillations at large values of
field or dot-size. Unlike laterally confined quantum dots, both (upper and lower) magneto-optical
transitions survive even in the extreme instances. However, the intra-Landau level transitions are
seen to be forbidden.
The spherical quantum dots have an edge over the strictly two-dimensional quantum dots in that the
additional (magnetic) quantum number makes the physics richer (but complex). A deeper grasp of the
Coulomb blockade, quantum coherence, and entanglement can lead to a better insight into promising
applications involving lasers, detectors, storage devices, and quantum computing.

\end{abstract}

\maketitle

\newpage

\section{Introduction}

The past two decades have seen very intense research efforts on the band-gap engineering in the synthetic
semiconducting systems. Thanks to the development of epitaxial-growth technology and electron lithography,
this field has created an excellent example of how the fundamental science and technology can interact
and influence one another. We refer, in particular, to the exotic physics emerging from the semiconducting
quasi-N-dimensional electron systems (QNDES); with $N$ [$\equiv$ 2, 1, or 0] being the degree of freedom.
While the seeds of the QNDES were sown much earlier [1], the discovery of quantum Hall effects is known to
have induced the titanic research interest to see the consequent changes with diminishing system's
dimensions from 3 to 2, 2 to 1, and 1 to 0. These are becoming known as quantum wells, quantum wires, and
quantum dots in which the charge carriers exposed to electric and/or magnetic fields can [and do] reveal
novel quantal effects that strongly modify the behavior characteristics of the resulting devices. The urge
for such miniaturization lies in the current quest for potentially {\em smaller} and {\em faster} devices.
A comprehensive review of the electronic, optical, and transport phenomena in the quantum structures of
reduced dimensionality can be found in Ref. [2].

What makes these QNDES so much attractive are the smaller effective masses of the charge carriers and
(substantially) large (background) dielectric constants. As a result, the charge carriers in
semiconductors are more responsive to minute compositional changes or small perturbations -- a situation
that is exploited for device applications and for making model systems unattainable with free particles.
The quantum size effects arise when the dimensions of a system become comparable to the de-Broglie
wavelength [$\lambda \sim$ 10 nm]. These are the crucial factors that spurred the fabrication of quantum
wells, wires, and dots with the same fundamental principle: confine the electrons in a semiconductor with
a smaller band-gap sandwiched between two identical semiconductors with a larger band-gap -- a measure of
the amount of energy needed to be pumped into the material to get electrons flowing. Thus an ultra thin
layer with free charge carriers forms the Q2D quantum well; a narrow strip sliced from the Q2D layer makes
a Q1D quantum wire; and dicing up a Q1D wire yields Q0D quantum dot.

By controlling the size and dimensions of such QNDES, researchers can tailor the electronic, optical,
and transport properties of a device at will. In theory, the fewer the dimension, the finer the tuning,
the sharper the clustering of energy states around specific peaks, and more important the many-body
effects. In addition, the reduced degrees of freedom allow detailed and often exact calculations. The
condensed matter physicists still fondly remember the excitement that followed the original proposals
for designing the quantum wells [3], quantum wires [4], and quantum dots [5]. If recent history (of
past 20 years) is any guide, it may not be an exaggeration to state that the vast majority of the
condensed matter physicists -- theorists and experimentalists -- are still overwhelmingly fascinated by
these low-dimensional, quantum structures. It seems to be true no matter if they name it the spintronics,
topological insulators, graphene, silicene, or germanene. The ability of these nanostructures to turn the
broad energy bands of the parent semiconductors into sharply defined energy levels is a transformation that
promises greater speed and efficiency for the resulting electronic and optical devices.

Ever increasing interest in these quantum structures is ascribed not only to their potential applications
but also to the fundamental physics involved [2]. For instance, the Q2DES (or quantum wells and
superlattices) allow the researchers a crystal clear observation of the Bloch oscillations, unobservable
in the conventional solids. The Q1DES (or quantum wires and lateral superlattices) offer us an excellent,
unique opportunity to study the real 1D Fermi gases in a relatively controlled manner. The Q0DES (or
quantum dots or antidots) allow us a testing ground for studying the electron-electron interactions in a,
neat and clean, tiny laboratory within the so-called artificial atoms.

Scientists have approached the fabrication of quantum dots from two very different points of view: (i) a
top-down approach in which the extent and dimensionality of the solid has gradually been reduced, and (ii)
a bottom-up in which quantum dots are viewed as extremely large molecules or colloids. The quantum dots
grown by epitaxial and lithographic techniques are in the size regime from 1 $\mu$m down to 10 nm, whereas
the colloidal samples vary in diameter from the truly molecular regime of 1 nm to about 20 nm. The latter
systems are also known in the literature by the names of colloidal quantum dots or nanocrystals. The
spherical quantum dots (SQDs), which are the subject matter of the present work, are precisely the systems
that belong to this family of systems and are playing better responsive role as lasers than their
epitaxial-cum-lithographic counterparts.

The research interest burgeoned in SQDs has focused mostly on the excitonic aspects of the bulk part with
the surface states, generally, eliminated by enclosure in a material of larger band-gap [6-35]. Existing
research work has devoted to understanding diverse fundamental properties of SQDs related with the exciton
dynamics. These include the distribution of electrons and holes, exciton formation, exciton binding energy,
exciton absorption, recombination, impact ionization, carrier multiplication, direct and indirect
transitions, exciton spin dynamics, optical non-linearity, oscillator strength, and excitonic atoms -- to
name a few -- in the theoretical [6-14, 17, 19, 21-22, 24-28, 30-35] and experimental [15, 16, 18, 20, 23,
29] works. Theoretical models employed for this purpose involve the tight-binding, effective-mass
approximation, variational approach, perturbative scheme, finite-element, pseudopotential, and the Kane's
${\bf k\cdot p}$ model in the systems of CdS, CdSe, PbTe/CdTe, and InAs/GaAs; and considering the
effect of an applied magnetic field [13, 17, 19-21, 34], the temperature dependence [23], the phonon
effects [25], the Coulomb impurities [30], and the surface effect [32].

Singular in the list of references is the Ref. 35 in which the authors have studied the SQDs made up of 3D
topological insulator (TI) materials, such as PbTe/Pb$_{1-x}$Sn$_{x}$Te, with bound massless and helical
Weyl states existing at the interface. The authors demonstrate a complete confinement of the massless Weyl
fermions at the interface and identify the spin locking and the Kramers degeneracy, the hallmarks of a 3D
TI. They argue that the semiclassical Faraday effect [due to the Pauli exclusion principle] can be used to
read out spin quantum memory in the optically mediated quantum computing. It is captivating given the
ongoing excitement behind the quantum computation and the advocacy of comparing the electronic, spintronic,
and optical processing of information.


The scrutiny of the existing literature on the spherical quantum dots reveals the lack of genuine efforts
devoted to theorizing the magneto-optical absorption that generally justifies the (localized) plasmon
excitation peaks observed in the optical experiments. The present paper is motivated to fill that gap. We
consider a SQD characterized by a confining harmonic potential and an applied magnetic field in the
symmetry gauge. We believe that the harmonic potential confining the SQDs is the most reasonable
approximation justifiable for the situation when the number of electrons ($N$) in the dots is small. This
model of harmonic potential validates the Kohn's theorem [36] -- generalized to Q0DES [37] -- which states
that the FIR resonant spectrum of a correlated many-electron system is insensitive to the interaction
effects. Identical physical conditions (of confinement and magnetic field) yield consistent results even
in relatively more sophisticated (spintronic) QNDES [38-40].

As such, we employ the Bohm-Pines' full-fledged random-phase approximation (RPA) [41] and derive
the total [or interacting] density-density correlation function (DDCF) in terms of the Dyson equation
that takes proper account of the Coulomb interactions. To this end, we make use of the eigenfunctions and
eigenenergies obtained as the solutions of the relevant Schrodinger equation in the spherical geometry. As
an application of the analytical results, we compute the Fock-Darwin spectrum [42-44], the magneto-optical
absorption, the Fermi energy, and the Radial distribution for electrons. The finiteness of the SQD is
accounted for by matching a proper boundary condition: the total eigenfunction vanishes at the surface of
the quantum dot. In the illustrative examples we stress upon the dependence on the dot-size, the confining
potential, and the magnetic field.

The rest of the paper is organized as follows. In Sec. II, we discuss the theoretical framework within the
RPA that allows us to derive the total DDCF in terms of the Dyson equation embodying properly the Coulomb
interactions, to devise the selection rules for the magneto-optical transitions, to compute the Fermi
energy, and to discuss the radial distribution of electrons. In Sec. III, we discuss several illustrative
examples of, e.g., the Fock-Darwin spectrum, Fermi energy, magneto-optical absorption, optical transitions,
and radial distribution of electrons. Finally, we conclude our findings and suggest a few interesting
features worth adding to the problem in Sec. IV.

\begin{figure}[htbp]
\includegraphics*[width=8.0cm,height=8.0cm]{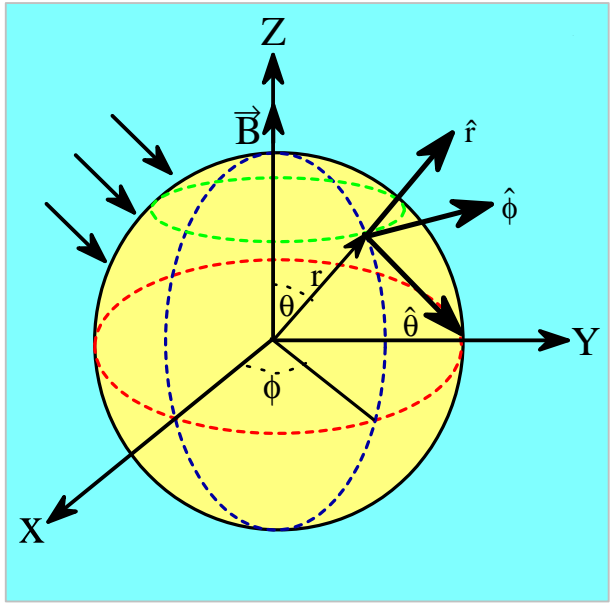}
\caption{(Color online) The schematics of the spherical geometry with coordinates ($r,\theta,\phi $):
Here $r$, $\theta$, and $\phi$ are, respectively, the radial coordinate, the polar angle, and the
azimuthal angle. The confinement potential $V$(${\bf r}$)$=\frac{1}{2}\,m^*\,\omega^2_o\,r^2$; and
the magnetic field in the symmetric gauge is defined by the vector potential
${\bf A}({\bf r})=\frac{1}{2}\,({\bf B}\times {\bf r})$.}
\label{fig1}
\end{figure}

\section{Theoretical framework}

\subsection{Eigenfunction and eigenenergy}

We consider a quasi-zero dimensional electron system three-dimensionally confined by a harmonic potential
$V({\bf r})=\frac{1}{2}\,m^*\,\omega_0^2\, r^2$ and subjected to an applied magnetic field in the
symmetric gauge [${\bf A}({\bf r})=\frac{1}{2}\,({\bf B}\times {\bf r})$] in the spherical geometry [with
${\bf r}\equiv (r, \theta, \phi)$; see Fig. 1]. For such a typical Q0DES in the spherical quantum dot, the
single-particle [of charge $-e$, with $e>0$] Hamiltonian can be expressed as
\begin{equation}
H = \frac{1}{2\,m^*}\, \Big [{\bf p} + \frac{e}{c}\,{\bf A}\Big ]^2 + \frac{1}{2}\,m^*\,\omega_o^2\, r^2\, ,
\end{equation}
Where $c$, $m^*$, ${\bf p}$, ${\bf A}$, ${\bf B}$, and $\omega_o$ are, respectively, the speed of light in
the vacuum, electron effective mass, momentum operator, vector potential, magnetic field, and the
characteristic frequency of the harmonic oscillator. Obviously, we consider a one-component plasma inside
the Q0DES and neglect the spin-orbit interactions and the Zeeman energy for the sake of simplicity. In the
light of the Coulomb gauge [$\nabla\cdot{\bf A}=0\Rightarrow {\bf A} \cdot {\bf p}={\bf p}\cdot {\bf A}$],
the Hamiltonian in Eq. (1) can be written as
\begin{equation}
H = - \frac{\hbar^2}{2\,m^*}\nabla^2 + \frac{1}{2}\frac{e\,B}{m^*\,c}\,\hat{L}_z +
     \frac{1}{8}\frac{e^2}{m^*c^2}\,({\bf B}\times {\bf r})^2 + \frac{1}{2}m^*\,\omega_o^2\, r^2\, ,
\end{equation}
where the operator $\hat{L}_z=-i\hbar \frac{\partial}{\partial \phi}$ is the z-component of the angular
momentum. Making use of the Laplacian operator $\nabla^2$ (in the spherical coordinates), substituting
$\Psi=R\Theta\Phi$ such that $\Psi(r,\theta,\phi)=R(r)\Theta(\theta)\Phi(\phi)$, and transposing allows
us to cast the Schrodinger equation $H\Psi=${\Large $\epsilon$}$\Psi$ -- {\Large $\epsilon$} being the
eigenenergy -- in the form
\begin{eqnarray}
\frac{1}{R}\frac{d}{dr}\Big (r^2\frac{dR}{dr} \Big) +
    \frac{2m^*r^2}{\hbar^2}\Big[\mbox {\Large $\epsilon$} -
        \frac{1}{2}\frac{eB}{m^*\,c}\hat{L}_z -
          \frac{1}{8}\frac{e^2}{m^*\,c^2}\,({\bf B}\times {\bf r})^2 -
            \frac{1}{2}m^*\,\omega_o^2\, r^2\Big] \nonumber\\
            + \frac{1}{\Theta \sin \theta}\frac{d}{d\theta}\Big(\sin \theta\frac{d\Theta}{d\theta}\Big)
              + \frac{1}{\Phi \sin^2\theta}\frac{d^2\Phi}{d\phi^2}=0\, .
\end{eqnarray}
What follows is the standard, lengthy but involved, mathematical procedure employing the method of
separation of variables and searching the step-wise solutions beginning first with the last term in
Eq. (3). The result is that the present system of Q0DES turns out to be characterized by the
eigenfunction
\begin{equation}
\Psi(r,\theta,\phi)=R_{nl}(r)\,Y^m_l(\theta,\, \phi)\, ,
\end{equation}
where the radial function
\begin{equation}
R_{nl}(r)=N_r \, e^{-\mbox{\scriptsize X}/2} \, \mbox{\scriptsize X}^{l/2} \,
             \Phi \big(-\alpha_{nl};\, 1+\mbox{\large s};\,\mbox{\scriptsize X}\big)\, ,
\end{equation}
where $\mbox{\large s}=\frac{1}{2}+l$, $\mbox{\scriptsize X}=r^2/l^2_H$, and $N_r$ is the normalization
coefficient defined by
\begin{equation}
N_r^{-2}=\frac{1}{2}\,l^3_H\,\int^{\overbar{\mbox{\scriptsize X}}}_0 d\mbox{\scriptsize X}\,
            e^{-\mbox{\scriptsize X}}\,
          \mbox{\scriptsize X}^{\mbox{\large s}}\,
         \big[\Phi\big(-\alpha_{nl};\, 1+\mbox{\large s};\, \mbox{\scriptsize X}\big)\big]^2\, ,
\end{equation}
where $\overbar{\mbox{\scriptsize X}}=\mbox{\scriptsize X}|_{r=R}$, and
$\Phi\big(-\alpha_{nl};\, 1+\mbox{\large s};\, \mbox{\scriptsize X}\big)$, $l_H=\sqrt{\hbar/(m^*\,\Omega_H)}$,
$\Omega_H=\sqrt{\frac{\omega^2_c}{4}\,\sigma^2+\omega^2_o}$, $\omega_c=eB/(m^*c)$, and $R$
are, respectively, the confluent hypergeometric function (CHF) [45], the hybrid magnetic length, the hybrid
characteristic frequency, the (electron) cyclotron frequency, and the dot radius. Figure 2 demonstrates that
$\Phi (...)$ is unambiguously a well-behaved function over a wide range of $X$ and $\alpha$ and that its period
is seen to be decreasing with increasing $X$ or $\alpha$ as the case may be. The spherical harmonics in Eq. (4)
are defined as [46]

\begin{figure}[htbp]
\includegraphics*[width=8.0cm,height=8.0cm]{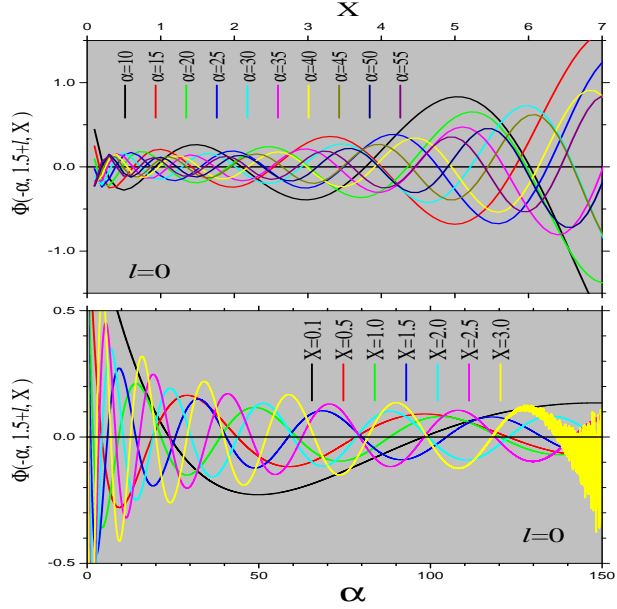}
\caption{(Color online) The plot of the confluent hypergeometric function (CHF)
$\Phi (\alpha_{nl}; \frac{3}{2}+l; X)$ vs. $X$ (the upper panel) and versus
$\alpha$ (the lower panel), for $l=0$. The purpose behind embedding this figure
was/is to demonstrate that the CHF is a well-behaved function over a wide range
of $X$ and $\alpha_{nl}$.}
\label{fig2}
\end{figure}

\begin{equation}
Y^m_l(\theta, \phi)=N_y\, P^m_l (\cos \theta)\, e^{i\,m\,\phi}\, ,
\end{equation}
where $P^m_l (\cos \theta)$ is the associated Legendre function and $N_y$ is the normalization coefficient
defined by
\begin{equation}
N_y=\Big[\frac{(2l+1)}{4\pi}\frac{(l-m)!}{(l+m)!} \Big]^{1/2}\, .
\end{equation}
In Eqs $(4)-(8)$ $n$, $l$, and $m$ are, respectively, the principal, orbital, and magnetic quantum numbers.
Remember that the spherical harmonics satisfy the orthonormality such as
\begin{equation}
\int d\Omega \,Y^{m*}_l(\theta, \phi)\,Y^{m'}_{l'}(\theta, \phi) = \delta_{ll'}\,\delta_{mm'}
\end{equation}
where the differential solid angle $d\Omega=\sin\theta\,d\theta\,d\phi$ and $\delta_{ij}$ is the well-known
Kronecker delta function. The finiteness of the quantum dot requires that the eigenfunction
$\Psi (...)$ satisfies the boundary condition $R_{nl} (r=R)=0\Rightarrow \Phi(-\alpha_{nl};\, 1+\mbox{\large s};\, \overbar{\mbox{\scriptsize X}})=0$. This determines the eigenenergy of the system defined by
\begin{equation}
\mbox {\Large $\epsilon$}_{nlm}=2\,\hbar\Omega_H\Big[\alpha_{nl}+
                 \frac{1}{2}\,\big(1+ \mbox{\large s}\big)\Big]+
                     \frac{1}{2}\,m\,\hbar\omega_c
\end{equation}
In the present form, the eigenenergy for the SQD is {\em formally} identical to that for the quantum dots
laterally confined in the Q2DES [2]. Before we close this section, we think that it is important to pinpoint
the appearance of a symbol $\sigma$ in the definition of $\Omega_H$ above. Once we arrive at solving the
radial part of the Schrodinger equation [see Eq. (3)], we realize that the third term in the square brackets
in Eq. (3) contains an angular term, namely $\sin^2\theta$. Of course, this is undesirable and therefore we
want to get rid of it. The best strategy is to take the matrix elements of the whole ({\em radial}) equation
between the states $|l,m>$ and $|l',m'>$. Doing this affects absolutely no other part of the radial equation
and converts this odd term into an expression defined by [47]
\begin{align}
\sigma^2
=& \Big<l'm'\Big|\sin^2\theta \Big|lm \Big>\nonumber\\
=& \delta_{l'l}\delta_{m'm}
 -\Big[\frac{(l+m+1)(l-m+1)}{(2l+1)(2l+3)}+\frac{(l+m)(l-m)}{(2l+1)(2l-1)}\Big]\delta_{l'l}\delta_{m'm}\nonumber\\
 &\,\,\,\,\,\,\,\,\,\,\,\,\,\,\,\,\,\,\,\,
 -\sqrt{\frac{(l+m)(l-m)(l+m-1)(l-m-1)}{(2l+1)(2l-1)^2(2l-3)}}\,\delta_{l',l-2}\delta_{m'm}\nonumber\\
 &\,\,\,\,\,\,\,\,\,\,\,\,\,\,\,\,\,\,\,\,
 -\sqrt{\frac{(l+m+2)(l-m+2)(l+m+1)(l-m+1)}{(2l+5)(2l+3)^2(2l+1)}}\,\delta_{l',l+2}\delta_{m'm}
\end{align}
This indicates that the magnetic field can induce the coupling between the states $l$ and $l\pm 2$. However,
this coupling is irrelevant to the magneto-optical transitions to be discussed in what follows. Another
important point is that the presence of the Kronecker delta functions [on both sides of the radial equation
while taking the said matrix elements] finally disallows the existence of the third and fourth terms in the
second equality of Eq. (11).

\subsection{Density-density correlation function}

The derivation of the density-density correlation function (DDCF) within the framework of Bohm-Pines' full-fledged
RPA is the main part of the theoretical framework. As it was designed, the RPA in which the electrons respond only
to the {\em total} (electric) potential accounts for the week screened Coulomb interactions and hence is widely
used for describing the nonlocal, dynamic, linear electronic response of the electron systems. To start with, we
recall the single-particle density-density correlation function (DDCF), which is given by [2]
\begin{equation}
\chi^o({\bf r}, {\bf r}';\omega)=\sum_{ij}\, \Lambda_{ij}\,
                            \Psi^*_i({\bf r})\Psi_j({\bf r})\Psi^*_j({\bf r}')\Psi_i({\bf r}')\, ,
\end{equation}
where
\begin{equation}
\Pi_{ij}=\frac{f(\epsilon_i)-f(\epsilon_j)}{\epsilon_i-\epsilon_j+\hbar\,\omega^+}\, ,
\end{equation}
where $f(\epsilon_j)$ is the standard Fermi-Dirac distribution function, $\Psi_j(...)$ is the single-particle eigenfunction [see Eq. (4)], and $\epsilon_j$ is the corresponding eigenenergy [see Eq. (10)]. The subindices
$i,j\equiv n,\,l,\,m$ and $\omega^+ =\omega+i\gamma$, where small but nonzero $\gamma$ refers to the adiabatic
switching of the Coulombic interactions in the remote past [48]. Next, we make use of the Kubo's correlation
function to write the induced particle density expressed as
\begin{align}
n_{in}({\bf r},\omega)
=&\int d{\bf r}\, \chi({\bf r}, {\bf r}';\omega)\,V_{ex}({\bf r}',\omega)\, , \\
=&\int d{\bf r}\, \chi^o({\bf r}, {\bf r}';\omega)\,V_{}({\bf r}',\omega)\, ,
\end{align}
where $V(...)=V_{ex}(...)+V_{in}(...)$ is the total potential, with suffix ex (in) referring to the external
(induced) potential in the system. Here $\chi (...)[\chi^o (...)]$ is the interacting [single-particle] DDCF.
Moreover, the induced potential is defined by
\begin{equation}
V_{in}({\bf r},\omega)=\int d{\bf r}\, V_{ee}({\bf r}, {\bf r}')\,n_{in}({\bf r}',\omega)\, ,
\end{equation}
where $V_{ee}(...)$ is the binary Coulombic interaction potential defined by
\begin{equation}
V_{ee}({\bf r}, {\bf r}')=\frac{e^2}{\epsilon_o}\frac{1}{\big|{\bf r}-{\bf r}'\big|}\, ,
\end{equation}
where $\epsilon_o$ is the dielectric constant of the background the Q0DES is embedded in. Now, it is not quite
difficult to derive, from Eqs. (14)-(17), the total [or interacting] DDCF that takes the form
\begin{equation}
\chi({\bf r}, {\bf r}';\omega)=\chi^o({\bf r}, {\bf r}';\omega) + \int d{\bf r}''\int d{\bf r}'''\,
                               \chi^o({\bf r}, {\bf r}'';\omega)\,V_{ee}({\bf r}'', {\bf r}''')\,
                                 \chi({\bf r}''', {\bf r}';\omega) \, .
\end{equation}
This is the famous Dyson equation in which the total DDCF $\chi(...)$ is, in general, a complex quantity. The
Dyson equation can be cast in the form
\begin{equation}
\int d{\bf r}''\,\epsilon({\bf r},{\bf r}'';\omega)\,\chi({\bf r}'',{\bf r}';\omega)=
              \chi^o({\bf r},{\bf r}';\omega)\, ,
\end{equation}
where the nonlocal, dynamic dielectric function $\epsilon({\bf r},{\bf r}';\omega)$ is defined as follows.
\begin{equation}
\epsilon({\bf r},{\bf r}';\omega)=\delta({\bf r}-{\bf r}')-
                                 \int d{\bf r}''\,\chi^o({\bf r},{\bf r}'';\omega)\,V_{ee}({\bf r}'',{\bf r}')\, .
\end{equation}
Multiplying on both sides of Eq. (19) by the inverse dielectric function
$\epsilon^{-1}({\bf r}''',{\bf r};\omega)$, integrating over ${\bf r}$, and making use of the identity:
\begin{equation}
\int d{\bf r}''\,\epsilon^{-1}({\bf r},{\bf r}'')\,\epsilon({\bf r}'',{\bf r}')=\delta({\bf r}-{\bf r}')\,
\end{equation}
enables us to rewrite Eq. (19) in the form
\begin{equation}
\chi({\bf r}, {\bf r}';\omega)= \int d{\bf r}''\,\chi^o({\bf r},{\bf r}'';\omega)\,
                                                 \epsilon^{-1}({\bf r}'',{\bf r}';\omega)\, .
\end{equation}
The inverse dielectric function $\epsilon^{-1}({\bf r},{\bf r}';\omega)$ was systematically derived for QNDES
(with $N\equiv$ 2, 1, 0) by Kushwaha and Garcia-Moliner [49] and is defined as follows.
\begin{equation}
\epsilon^{-1}({\bf r},{\bf r}';\omega)=\delta({\bf r}-{\bf r}')+\sum_{\mu\nu}\,
                                       L^*_{\mu}({\bf r})\,\Pi_{\mu}\,\Lambda_{\nu\mu}\,S_{\nu}({\bf r}')\, ,
\end{equation}
where the symbols $L_{\mu}({\bf r})$ and $S_{\mu}({\bf r})$ representing, respectively, the long-range and the
short-range parts of the response function are defined by
\begin{align}
L_{\mu}({\bf r})&=L_{ij}({\bf r})=\int d{\bf r}''\,
                        \Psi_i({\bf r}'')\,V_{ee}({\bf r}, {\bf r}'')\,\Psi^*_j({\bf r}'')\, ,\\
S_{\mu}({\bf r})&=S_{ij}({\bf r})=\Psi^*_j({\bf r})\,\Psi_i({\bf r})\, ,
\end{align}
and $\Lambda_{\mu\nu}$ is the inverse of $[\delta_{\mu\nu}-\Pi_{\mu}\beta_{\mu\nu}]$ such that
\begin{equation}
\sum_{\mu}\,\Lambda_{\gamma\mu}\,[\delta_{\mu\nu}-\Pi_{\mu}\beta_{\mu\nu}]=\delta_{\gamma\nu}\, ,
\end{equation}
where the symbol $\beta_{\mu\nu}$ stands for
\begin{equation}
\beta_{\mu\nu}=\int d{\bf r}\, L^*_{\mu}({\bf r})\,S_{\nu}({\bf r})
\end{equation}
Notice that the subscript $\mu\equiv i,\,\, j$ is a composite index introduced just for the sake of the
mathematical convenience [49]. Equation (22), with the aid of Eq. (23), can be rigorously expressed --
[after suppressing the $\omega-$dependence for the sake of brevity] -- as
\begin{equation}
\chi({\bf r},{\bf r}';\omega)=\sum_{ijkl}\,\Big[\Pi_{ij}\,\delta_{ik}\,\delta_{jl}+\Pi_{ij}\,\sum_{mn}\,
                       \Lambda_{klmn}\,\Pi_{mn}\,F_{mnij}\Big]\,
                \Psi^*_i({\bf r})\Psi_j({\bf r})\Psi^*_l({\bf r}')\Psi_k({\bf r}')\, ,
\end{equation}
where the symbol $F_{ijkl}$ stands for the matrix elements of Coulombic interactions and is expressed as
\begin{equation}
F_{mnij}=\int d{\bf r}''\int d{\bf r}'''\,
                  \Psi^*_m({\bf r}'')\Psi_n({\bf r}'')\,V_{ee}({\bf r}'',{\bf r}''')\,
                  \Psi^*_j({\bf r}''')\Psi_i({\bf r}''')
\end{equation}
In order to investigate the magneto-optical absorption in the semiconducting SQDs, what we actually need
to compute sagaciously is only the energy dependence of the {\em imaginary} part of $\chi (...)$, where
\begin{align}
\chi(\omega)
=&\int d{\bf r}\int d{\bf r}'\,\chi({\bf r},{\bf r}';\omega)\nonumber\\
=& \sum_{ijkl}\,\Big[\Pi_{ij}\,\delta_{ik}\,\delta_{jl}+\Pi_{ij}\,\sum_{mn}\,
                      \Lambda_{klmn}\,\Pi_{mn}\,F_{mnij}\Big]\,\nonumber\\
 & \hspace {3.25cm} \times \int d{\bf r}\int d{\bf r}'\,
                      \Psi^*_i({\bf r})\Psi_j({\bf r})\Psi^*_l({\bf r}')\Psi_k({\bf r}')\, .
\end{align}
The only subtlety left out heretofore is the use of the Laplace expansion, which is, in fact,the expansion
of the inverse distance between two points such as it occurs in the Coulomb potential. Let the points have
position vectors ${\bf r}$ and ${\bf r}'$, then the Laplace expansion is [50]
\begin{equation}
\frac{1}{\big|{\bf r}-{\bf r}'\big|}=\sum^\infty_{l=0}\,\sum^l_{m=-l}\,
       \frac{4\pi}{2l+1}\,\frac{r^l_<}{r^{l+1}_>}\,Y^{m*}_l(\theta, \phi)\,Y^{m}_{l}(\theta', \phi')\, ,
\end{equation}
where $r_<$ is the min($r, r'$) and $r_>$ is the max($r, r'$). Three simple steps -- (i) write the inverse
distance in the scalar form, (ii) deploy the generating function for the Legendre polynomial, and (iii)
use the spherical harmonic addition theorem -- allow the reader to derive the Laplace expansion in no time.

\subsection{The optical transitions: Selection rules}

In a completely confined system (as is the case here), the complex energy spectrum generally brings about
complicated structures in the transport characteristics. The optical transitions, on the other hand, follow
selection rules and are simpler to analyze. Here we examine the interaction between an applied oscillating
electric field E and the electric dipole moments of the electrons. The transition probability between the
electronic states $\big|n,l,m\big>$ and $\big|n',l',m'\big>$ in the SQDs, according to Fermi's golden rule,
is given by
\begin{equation}
M(n,l,m; n',l',m')=\frac{\pi}{2\hbar}\,
               \Big|\big<n',l',m'\big|r\,\sin\theta\, e^{\pm i\phi} \big|n,l,m \big>\Big|^2\, ,
\end{equation}
whereas the transition amplitude is expressed as [47]
\begin{align}
A(n,l,m; n',l',m')
&=\Big<n',l',m'\big|r\,\sin\theta\, e^{\pm i\phi} \big|n,l,m \Big>\nonumber\\
&=\int d{\bf r}\,R^*_{n'l'}(r)\,Y^{m'*}_{l'}(\theta,\phi)\,r\,\sin\theta\,e^{\pm i \phi}\,
                 R_{nl}(r)\,Y^m_l(\theta, \phi)\nonumber\\
&=\int dr\,r^3\,R^*_{n'l'}(r)\,R_{nl}(r)\,\int d\Omega\,
            Y^{m'*}_{l'}(\theta,\phi)\,\sin\theta\,e^{\pm i \phi}\,Y^m_l(\theta, \phi)\nonumber\\
&=P(l,m)\, \int dr\,r^3\,R^*_{n'l'}(r)\,R_{nl}(r)\, ,
\end{align}
where the symbol $P(l,m)$ is defined as

\begin{align}
P(l,m)=\left \{
\begin{array}{l}
+\Big[\sqrt{\frac{(l+m+1)(l+m+2)}{(2l+1)(2l+3)}}\,\delta_{l',l+1} -
\sqrt{\frac{(l-m)(l-m-1)}{(2l-1)(2l13)}}\,\delta_{l',l-1}\Big]\,\delta_{m',m+1}\, ; {\rm \,\,for\,\, +\,\, sign}\\
-\Big[\sqrt{\frac{(l-m+1)(l-m+2)}{(2l+1)(2l+3)}}\,\delta_{l',l+1} -
\sqrt{\frac{(l+m)(l+m-1)}{(2l-1)(2l13)}}\,\delta_{l',l-1}\Big]\,\delta_{m',m-1}\, ; {\rm \,\,for\,\, -\,\, sign}
\end{array}
\right.
\end{align}
Thus $P(l,m)$ already defines the selection rules for the orbital and the magnetic quantum numbers. Equation (33),
with the aid of Eq. (5), assumes the following form:
\begin{align}
A(n,l,m; n',l',m')
=&\frac{1}{2}\,P(l,m)\,N_r(n'l')\,N_r(nl)\,l^4_H\,\nonumber\\
 &\times  \int d\mbox{\scriptsize X}\,e^{-\mbox{\scriptsize X}}\,
                         \mbox{\scriptsize X}^{(\mbox{\large s}'+\mbox{\large s}+1)/2}\,
           \Phi\big(-\alpha_{n'l'};\, 1+\mbox{\large s}';\, \mbox{\scriptsize X}\big)\;\,
           \Phi\big(-\alpha_{nl};\, 1+\mbox{\large s};\, \mbox{\scriptsize X}\big)\, .
\end{align}
Since our main goal now is to determine the correlation between $n'$ and $n$, it does not matter if we, for the
moment, replace $\alpha_{nl}$ by $n$. Let us recall the identity relating the associated Laguerre polynomial
$L^{\mbox{\large s}}_n (\mbox{\scriptsize X})$ with the confluent hypergeometric function
$\Phi (-n;\, 1+\mbox{\large s};\, \mbox{\scriptsize X})$:
\begin{equation}
L^{\mbox{\large s}}_{n}(\mbox{\scriptsize X})=\binom{n+\mbox{\large s}}{n}\,
                                  \Phi(-n;\, 1+\mbox{\large s};\, \mbox{\scriptsize X})\, .
\end{equation}
The integral in Eq. (35), with the aid of Eq. (36), takes the form
\begin{align}
I
=&{\binom {n'+\mbox{\large s}'}{n'}}^{-1}\,{\binom {n+\mbox{\large s}}{n}}^{-1}\,
           \int d\mbox{\scriptsize X}\, e^{-\mbox{\scriptsize X}}\,
             \mbox{\scriptsize X}^{(\mbox{\large s}'+\mbox{\large s}+1)/2}\,
               L^{\mbox{\large s}'}_{n'}(\mbox{\scriptsize X})\,L^{\mbox{\large s}}_{n}(\mbox{\scriptsize X})\\
=&\left \{
\begin{array}{l}
      \frac{n!\,(\mbox{\large s}+1)!\,\,\mbox{\large s}!}{(n+s)!}\,\,\Big[\delta_{n',n} - \delta_{n',n-1}\Big]
                             ;\,\,\,\,\, {\rm \,\,\mbox{\large s}'=\mbox{\large s}+1}\\
      \frac{n'!\,\mbox{\large s}!\,(\mbox{\large s}-1)!}{(n'+s)!}\,\Big[\delta_{n',n} - \delta_{n',n+1}\Big]
                             ;\,\,\,\,\, {\rm \,\,\mbox{\large s}'=\mbox{\large s}-1}\, .
\end{array}
\right.
\end{align}
Therefore the selection rules that follow from Eqs. (34) and (38) are: $\Delta n=0,\,\, \pm 1$, $\Delta l=\pm 1$
and $\Delta m=\pm 1$. The corresponding transition energies are therefore specified by
\begin{equation}
\Delta{\mbox{\Large $\epsilon$}}=2\,\Big[\Delta\alpha\,\pm\,\frac{1}{2}\Big]\,\hbar\Omega_H\,
                                      \pm \,\frac{1}{2}\,\hbar\omega_c
\end{equation}
where $\Delta\alpha\equiv \Delta\alpha_{nl}=\big[\alpha_{1,1}-\alpha_{1,0}\big]$. As will be seen in the next
section, the transition energy plotted against the magnetic field reveals some interesting features. For the
zero confinement potential [i.e., $\hbar\omega_o=0$], the transition is purely cyclotron resonance and the
slope of the curve should define the electron effective mass $m^*$ in a typical manner due, in particular, to
the occurrence of the parameter $\sigma$ and the term $\Delta\alpha$. For the zero magnetic field [i.e., $B=0$],
the transition will also differ from the case of $\Delta ${\Large $\epsilon$}$=\hbar\omega_o$, again due to the
presence of the term $\Delta\alpha$. Only in the case of laterally confined quantum dots do the corresponding
cases become the first guess [2].

\subsection{The Fermi energy}

The Fermi energy is one of the most important concepts in the physics of solids as well as of quantum liquids
such as the low-temperature liquid He (both normal and superfluid). Although, it is a single-particle aspect,
it dictates largely the electron dynamics that is paramount to the understanding of electronic, optical, and
transport phenomena in fermionic systems.
Confusingly, the term Fermi energy is often used to refer to a
different but closely related concept: the Fermi level -- also called chemical
potential. A few fundamental differences between the Fermi energy and the Fermi level are as follows: (i) the
Fermi energy is only defined at absolute zero, while the Fermi level is defined for any temperature, (ii) the
Fermi energy is an energy difference, whereas the Fermi level is a total energy level (including the kinetic
and potential energies), and (iii) the Fermi energy can only be defined for non-interacting fermions, whereas
the Fermi level remains well defined even in complex interacting systems.

Since each quantum level takes two electrons with opposite spin, the Fermi energy $\mbox {\Large $\epsilon$}_F$
of a system of N non-interacting electrons at absolute zero is actually the energy of the ($\frac{1}{2}N$)th
level. In the case of a quantum dot containing $N$ electrons -- with complete confinement -- the Fermi energy
can be computed self-consistently through the following expression:
\begin{equation}
N = 2\, \sum_{nlm}\, \theta(\mbox{\Large $\epsilon$}_F-\mbox{\Large $\epsilon$}_{nlm})\, ,
\end{equation}
where the $\mbox{\Large $\epsilon$}_{nlm}$ is as defined in Eq. (10). In a completely confined system of quantum
dots, the single-particle excitation spectrum [see Sec. III] turns out to be quite complex and therefore the
Fermi energy cannot be expected to be a smooth function of the magnetic field. The most striking characteristic
of the semiconducting SQDs is the size-dependence of the Fermi level brought about by the variations in the
density of electronic states due, in fact, to the framed bounds of the nanocrystal. This can be understood by
recalling the Heisenberg's uncertainty principle in relation between position and momentum in free and confined
particles. For a free particle, or a particle in the periodic potential of an extended crystal, the momentum
can be precisely defined, but the position cannot. For a confined particle, while the uncertainty in the
position decreases, the momentum is no longer well-defined. The result is that the close by transitions occurring
at slightly different energies in an extended crystal will be squeezed by the quantum confinement into a single,
energetic transition in a quantum dot.

\section{Illustrative examples}

In order to discuss the illustrative numerical examples, it is indispensable to specify the material the quantum
dots are made of and the parameters used in the computation. We consider colloidally prepared GaAs spherical
quantum dots, which implies that the background dielectric constant $\epsilon_o=12.8$ and the effective mass of
the electron $m^*=0.067\,m_o$, where $m_o$ is the mass of the bare electron. The other parameters involved in
the computation are: the radius of the quantum dot ($R$), the confinement potential $\hbar\omega_o$, and the
magnetic field $B$; which will be stated while discussing the specific results for the respective cases. It is
also imperative to limit the number of electrons -- or, in other words, to specify the quantum numbers $n$, $l$,
and $m$ -- while computing the optical (or magneto-optical) absorption in the system [see, e.g., Eqs. (28) or
(30)]. This is equivalent to truncating the $\infty\times\infty$ matrix in the band structure computation in
solid state physics by limiting, for example, the number of plane waves. Until and unless stated otherwise, we
have specified the quantum numbers such that the quantum dot contains in total {\em twenty four} electrons
(including spin). It is noteworthy that we also assume the compliance of the lowest subband approximation due to
the harmonic confinement potential, which makes sense for the quantum dots with small charge densities at low
temperatures where most of the experiments are performed on the low-dimensional systems. For the sake of
comparison, we also discuss briefly the results for the case of zero magnetic field.

\begin{figure}[htbp]
\includegraphics*[width=8.0cm,height=8.0cm]{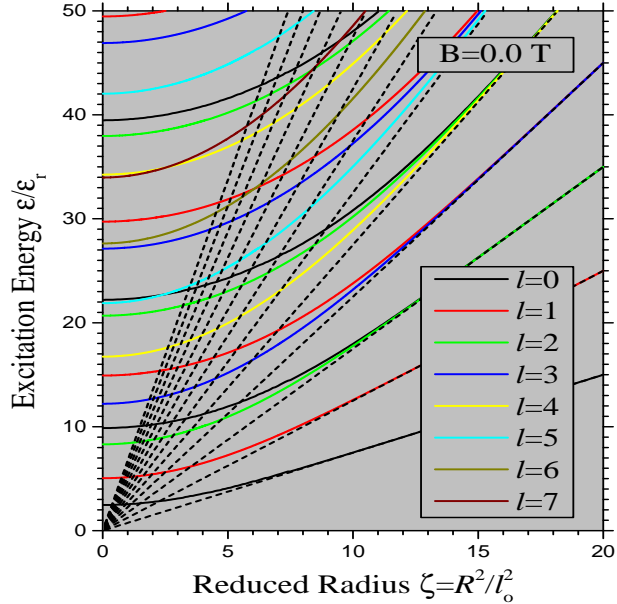}
\caption{(Color online) The single-particle excitation spectrum for the GaAs SQDs in the absence of an
applied magnetic field, i.e. $B=0$, for several values of the orbital quantum number $l$. The electric
fan plotted as dashed lines is a zero magnetic-field analogue of the magnetic (or Landau) fan. The
symbols are as defined in the text.}
\label{fig3}
\end{figure}

\subsection{The zero magnetic field:}

Figure 3 illustrates the single-particle excitation spectrum for the harmonically confined SQDs in the absence
of an applied magnetic field. The plots are rendered in terms of the dimensionless variables:
$\epsilon_{nl}/\epsilon_r$ and $\zeta=R^2/l^2_o$. Here $\epsilon_r=2\,\hbar^2/(m^*\,R^2)$, and $l_o=\sqrt{\hbar/(m^*\omega_o)}$ is the characteristic length due to the harmonic confinement. Most of the
modes start from zero with close to zero group velocity, which becomes gradually positive as the excitation
energy increases with increasing $\zeta$. At large values of $\zeta$, there is a set of $j$ (with $j\ge 2$)
branches -- belonging to different quantum numbers $n$ and $l$ -- that are seen to merge together. We find
that one such set can be generated by $\sum^n_{k=1}\,[n-(k-1), 2(k-1)]$ and the other by
$\sum^n_{k=1}\,[n-(k-1), (2k-1)]$, with $n\ge 2$. Thus, the coalesced branches in terms of ($n, l$) are:
(2, 0)$+$(1, 2); (3, 0)$+$(2, 2)$+$(1, 4); (2, 1)$+$(1, 3); and (3, 1)$+$(2, 3)$+$(1, 5). The dashed lines
represent the {\em electric} fan due to the confinement given by $\epsilon/\epsilon_r=\zeta\,$[$l+\frac{3}{2}$]
-- analogous to the {\em magnetic} (or Landau) fan for the non-zero magnetic field. For very large confinement
potential, we obtain the ideal 3D limit defined by $\epsilon_{nl}=\frac{3}{2}\hbar\omega_o$. We believe that
the transport measurements on SQDs should be capable to map the excitation energy levels shown in Fig. 3

\begin{figure}[htbp]
\includegraphics*[width=8.0cm,height=8.0cm]{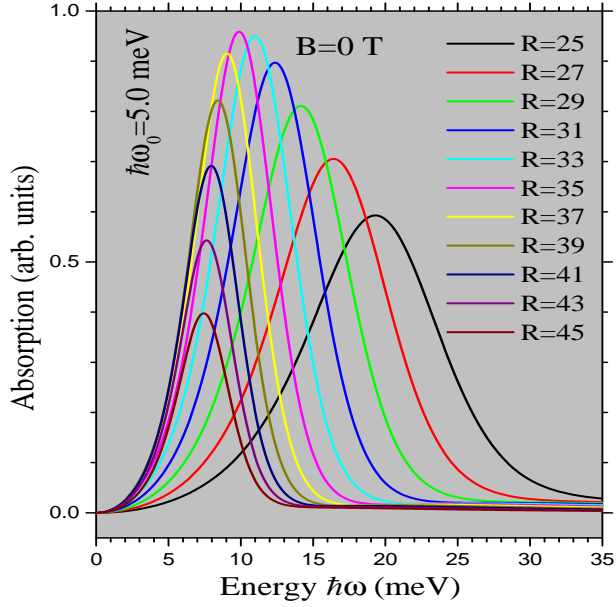}
\caption{(Color online) The optical absorption vs. the excitation energy for the GaAs SQDs in the
absence of an applied magnetic field, for several values of the dot radius $R$. The confinement
potential $\hbar\omega_o=5$ meV.}
\label{fig4}
\end{figure}

Figure 4 shows the optical absorption, i.e., Im$\chi (\omega)$ [see Eq. (30)], versus the excitation energy for
a given confinement potential $\hbar\omega_o=5$ meV and for several values of the dot radius. We consider the
quantum dots with radii lying in the range defined by $25 \le R\, (\rm{nm})\le 45$. This size range is known to
be quite easily attainable in the colloidally prepared SQDs. Note that we have scaled down the vertical axis
with no loss of generality. What we observe is that as the size of the quantum dot is reduced, the absorption
peak undergoes a blue shift (in energy). In other words, diminishing the dot-size pushes the ({\em localized})
plasmon excitations to shift to higher energy, thereby restricting the oscillator strength to be concentrated
into just a few transitions. The quantum confinement brings about these physical occurrences due to the changes
in the electronic density of states. This implies that tailoring the size of the quantum dots leads us to tune
the electronic excitations at will.

\begin{figure}[htbp]
\includegraphics*[width=8.0cm,height=8.0cm]{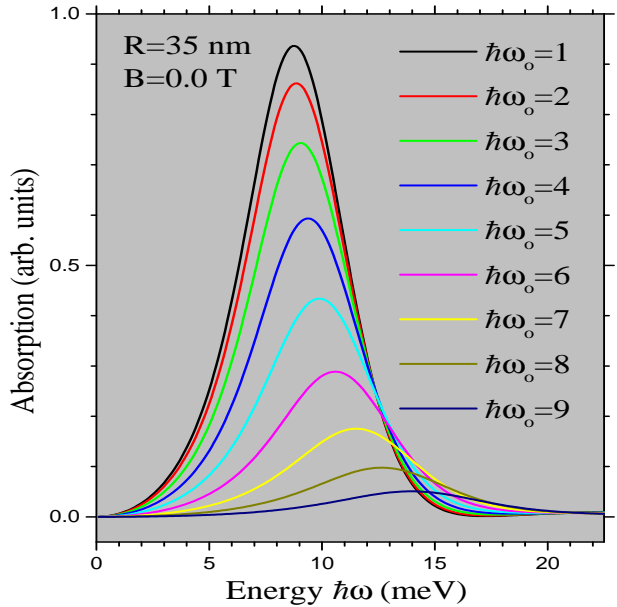}
\caption{(Color online) The optical absorption vs. the excitation energy for the GaAs SQDs in the absence
of an applied magnetic field, for several values of the confinement potential $\hbar\omega_o$. The dot
radius $R=35$ nm.}
\label{fig5}
\end{figure}

Figure 5 depicts the optical absorption as a function of the excitation energy for a given dot radius $R=35$ nm
and for several values of the confinement potential ($1\le \hbar\omega_o$ (meV) $\le 9$). We observe that the
absorption peak height (width) reduces (broadens) with increasing confinement potential. It is important to
notice that the peak position observes a blue shift as the confinement potential grows. To be specific, we
notice that if the confinement grows by a factor of nine, the peak position experiences a blue-shift of $60\%$.
This is understandable: the stronger the confinement potential, the smaller the space available to the electrons
(or greater the degree of localization). This implies virtually the reduction in the dot size and hence the blue
shift of the absorption peaks.

\begin{figure}[htbp]
\includegraphics*[width=8.0cm,height=8.0cm]{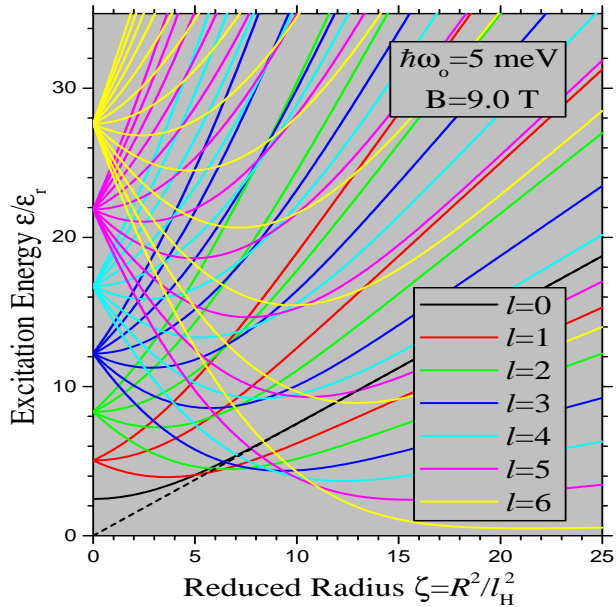}
\caption{(Color online) The single-particle excitation spectrum for the GaAs SQDs in the presence of an
applied magnetic field ($B$), for several values of the orbital quantum number $l\le 6$. The dashed line
represents the lowest branch of the magnetic (or Landau) fan -- when $\alpha_{nl}\rightarrow 0$ for very
large $R$ or $B$. The magnetic field $B=9$ T and the confinement potential $\hbar\omega_o=5$ meV.}
\label{fig4}
\end{figure}

\subsection{The non-zero magnetic field:}

Figure 6 represents the single-particle excitation spectrum for the spherical quantum dot in the presence of a
confinement potential ($\hbar\omega_o=5$ meV) and an applied magnetic field ($B=9$ T) for the orbital quantum
number $l\le 6$. The plots are rendered in terms of the dimensionless energy $\epsilon_{nlm}/\epsilon_r$ and
the dot radius $\zeta=R^2/l^2_H$. Here $l_H=\sqrt{\hbar/(m^*\Omega_H)}$ is the hybrid magnetic length [see Sec.
II.A] and $\epsilon_r$ is just as defined above. First and foremost, we observe that, unlike the laterally
confined 2D quantum dots [2], the electronic levels at the origin [i.e., at $\zeta=0$] are not equispaced.
This is clearly attributed to the geometrical difference more than anything else. This remark is also valid
in relation with Fig. 3. The ($2\,l+1$)-degeneracy at the origin justifies the intuition. All the modes with
$m\ge 0$ are seen to start and propagate throughout with a positive group velocity, whereas those with $m<0$
start with a negative group velocity, attain a minimum as $\zeta$ increases, and finally propagate with the
positive group velocity. It is noticeable that the larger the $|m|$, the greater the value of $\zeta$ where
the latter type of modes observe the minimum. The dashed line starting from zero and finally merging with
($l=0=m$) mode -- the ground state -- is the lowest branch of the magnetic [or Landau] fan. We did not plot
the whole magnetic fan in order not to make a mess inside the figure. At large values of magnetic field --
keeping the dot radius and the confinement potential fixed -- $\zeta$ becomes large, the role of $\alpha_{nl}$
diminishes and the magnetic fan is born. In case the confinement energy predominates over the cyclotron energy
(i.e., $\hbar\omega_o \gg \hbar \omega_c$), Fig. 3 is reinstated.

\begin{figure}[htbp]
\includegraphics*[width=8.0cm,height=8.0cm]{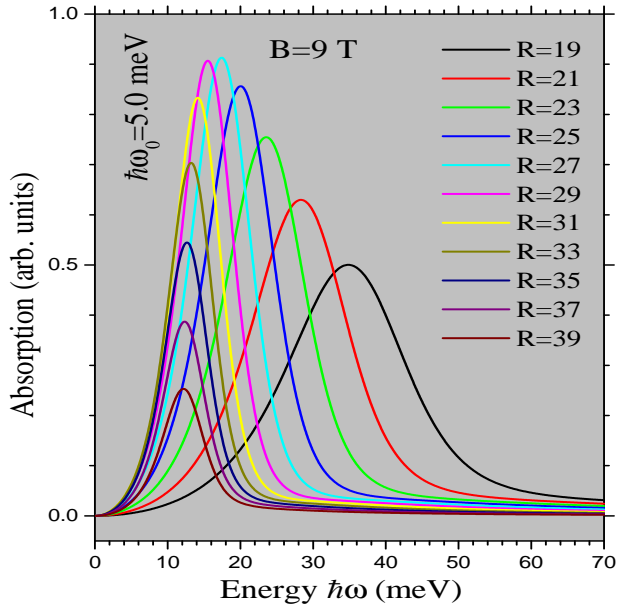}
\caption{(Color online) The magneto-optical absorption vs. the excitation energy for the GaAs SQDs in the
presence of an applied magnetic field ($B$), for several values of the dot radius $R$. The magnetic field
$B=9$ T and the confinement potential $\hbar\omega_o=5$ meV.}
\label{fig7}
\end{figure}

Figure 7 portrays the magneto-optical absorption as a function of the excitation energy for a given value of
the confinement potential ($\hbar\omega_o=5$ meV) and the magnetic field ($B=9.0$ T), for several values of
the dot radius ($19\le R\,(\rm{nm})\le 39$). It is observed that the absorption peak experiences a blue shift
as the dot radius is decreased. A similar observation was made in the case of a zero magnetic field
[see, e.g., Fig. 4]. Comparing Fig. 7 with Fig. 4 leads us to notice that the magnetic field enhances the blue
shift for a given dot size. In other words, the magnetic field prompts the quantum dot to absorb photons with
higher energies. For example, for $R=35$ nm [$R=39$ nm] the energy peak has blue-shifted by $31\%$ [$46\%$].
This also indicates that the larger the dot size, the greater the effect of the magnetic field. This seems to
make sense. In a very constricted dot, the charge carriers do not have sufficient room to respond to and hence
to feel the effect of an applied magnetic field. It has also been noticed that the FWHM of the absorption peak
increases with decreasing dot size. 

\begin{figure}[htbp]
\includegraphics*[width=8.0cm,height=8.0cm]{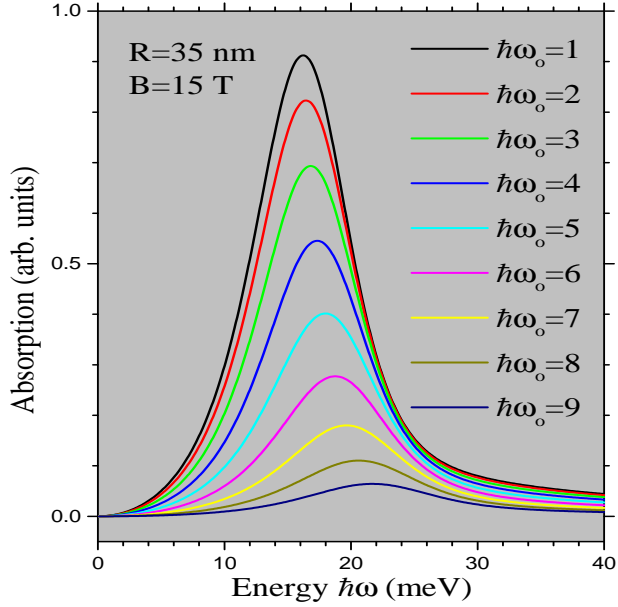}
\caption{(Color online) The magneto-optical absorption vs. the excitation energy for the GaAs SQDs in the
presence of an applied magnetic field ($B$), for several values of the confinement potential $\hbar\omega_o$.
The magnetic field $B=15$ T and the dot radius $R=35$ nm.}
\label{fig8}
\end{figure}

Figure 8 sketches the magneto-optical absorption versus the excitation energy for a given value of the dot size
($R=35$ nm) and the magnetic field ($B=15$ T), for several values of the confinement potential
($1\le \hbar\omega_o$ (meV) $\le 9$). One observation -- common to Fig. 7 -- is that the absorption peak is
blue-shifted as the confinement potential increases. A similar effect was observed in relation with Fig. 5 for
the zero magnetic field. To be specific, we notice that for $\hbar\omega_o=1$, 5, 9 meV the absorption peak has
blue-shifted, respectively, by $84\%$, $82\%$, and $55\%$. This also shows that the stronger the confinement,
the weaker the blue-shift experienced by the absorption peaks. Again, the FWHM of the absorption peak increases
with increasing confinement. This also reflects the tendency of the oscillator strength focused into a fewer
transitions for the stronger confinement. 

\begin{figure}[htbp]
\includegraphics*[width=8.0cm,height=8.0cm]{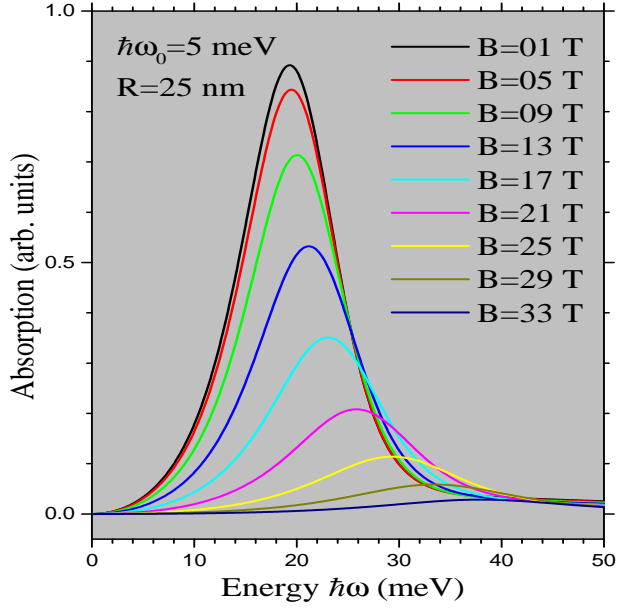}
\caption{(Color online) The magneto-optical absorption vs. the excitation energy for the GaAs SQDs in the
presence of an applied magnetic field ($B$), for several values of the magnetic field $B$. The dot radius
$R=25$ nm and the confinement potential $\hbar\omega_o=5$ meV.}
\label{fig9}
\end{figure}

Figure 9 renders the magneto-optical absorption against the excitation energy for a given value of the dot size
($R=25$ nm) and the confinement potential ($\hbar\omega_o=5$ meV), for several values of the applied magnetic
field ($1\le B$ (T) $\le 33$). A few first-hand observations are: the peak height decreases, FWHM increases,
and the blue-shift is enhanced with increasing intensity of the applied magnetic field. To be a little bit more
specific, we observe that increasing the magnetic field from $B=1$ T to $B=25$ T results in the absorption
peak's blue-shift by $52\%$ for the present set of parameters. This defines the role of an applied magnetic
field in the optical absorption of spherical quantum dots. That is the stronger the magnetic field, the greater
the capacity of the quantum dots to absorb the photons of higher energy. The other remarks regarding the
oscillator strength about Fig. 8 are still valid. By and large, the size effects predominate over those due to
confinement and magnetic field.

\begin{figure}[htbp]
\includegraphics*[width=8.0cm,height=9.0cm]{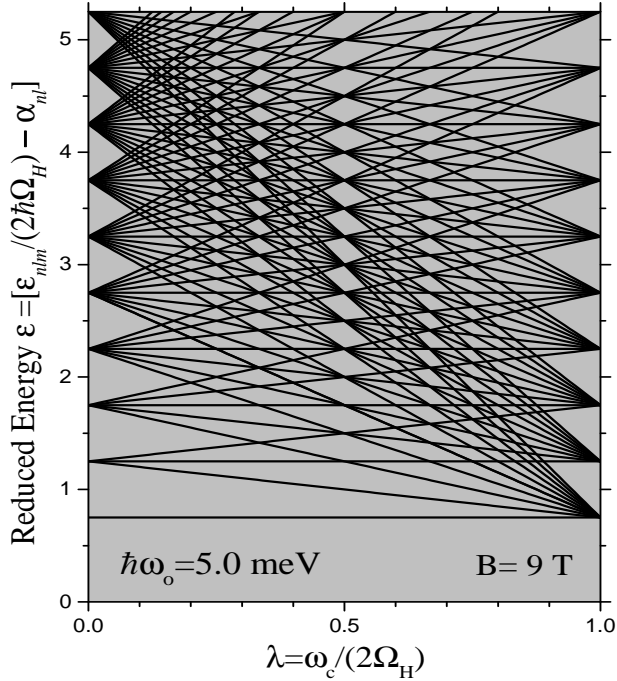}
\caption{(Color online) The zigzag representation of the single-particle excitation spectrum for the GaAs SQDs
for the confinement potential $\hbar\omega_o=5$ meV and the magnetic field $B=9$ T [see the text for the
expression of $\epsilon$]. Notice that following the state ($l = 1$), crossings with states of ($l < 0$) occur
at $\lambda=(l - 1)/(l + 1)$.}
\label{fig10}
\end{figure}

\subsection{Other field-effects in SQDs}

Figure 10 draws the zigzag type of single-particle excitation spectrum for SQDs in the presence of an applied
magnetic field ($B=9$ T) and the confinement potential ($\hbar\omega_o=5$ meV). Specifically, we plot the
dimensionless energy $\epsilon$ [$=\epsilon_{nlm}/(2\hbar\Omega_H)-\alpha_{nl}$]
$\,= \frac{1}{2}(\frac{3}{2}+l)+\frac{1}{2}\,m\,\lambda$ versus the dimensionless parameter $\lambda=\omega_c/(2\,\Omega_H)$ [see Eq. (10)]. The interest in such eigenenergies as plotted in this figure
lies in the fact that they allow us to debate the gradual transitions from pure spatial quantization
[$\omega_o\ne 0$, $\omega_c=0$, $\lambda=0$] to pure magnetic quantization
[$\omega_o= 0$, $\omega_c\ne 0$, $\lambda=1$]. In this representation each state $\psi_{nlm}$ corresponds to
one straight line with slope given by $\frac{1}{2}\, m$. For $m\le 0$, the quantum number $l$ dictates the
crisscrossing with the vertical axis $\lambda=1$. At $\lambda=0$, the ($2\,l+1$)-degeneracy is given by the
$\epsilon$, whereas at $\lambda=1$, we acquire the familiar macroscopic degeneracy of magnetic (or Landau)
levels. Note that we have limited to states with orbital quantum number $l\le 9$, otherwise there would be
innumerable number of lines of increasing slope emerging from each of the levels at $\lambda=1$. Obviously,
only ($l=0=m$) states are independent of $\lambda$ and constitute the straight lines with zero slope.
Following the state ($l=1$), crossings with states of ($l < 0$) occur at $\lambda=(l-1)/(l+1)$.

\begin{figure}[htbp]
\includegraphics*[width=8.0cm,height=9.0cm]{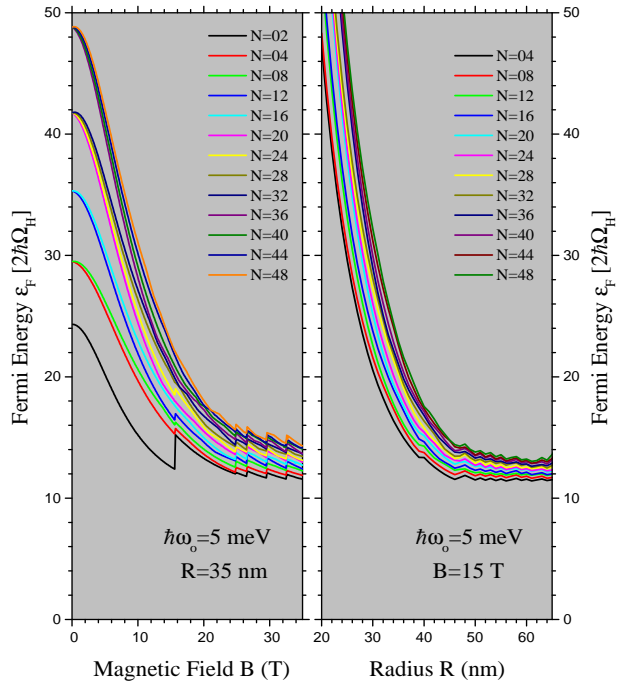}
\caption{(Color online) The Fermi energy of the N-particle GaAs spherical quantum dot as a function of the
magnetic field ($B$) [the left panel] and the dot radius ($R$) [the right panel], for the confinement
potential $\hbar\omega_o=5$ meV. The dot radius $R=35$ nm [in the left panel] and the magnetic field $B=15$ T
[in the right panel].}
\label{fig11}
\end{figure}

Figure 11 presents the dot-size and magnetic-field dependence of the Fermi energy, self-consistently computed
through Eq. (40), for several numbers of the electrons comprising the quantum dot. The left (right) panel
shows the magnetic-field (dost-size) dependence of the Fermi energy for a given value of the confinement
potential ($\hbar\omega_o=5$ meV) for $4\le N \le 48$. We fix the dot-size (magnetic field) to be $R=35$ nm
($B=15$ T) in the left (right) panel. It is observed that the Fermi energy gradually decreases with increasing
magnetic field or the dot-size. At large magnetic fields, the Fermi energy is seen to be making saw-tooth-like
oscillations with increasing $B$. The similar, but less pronounced, behavior is noticed at larger dot-size in
the right panel. However, the larger the number of electrons, the greater the Fermi energy for all values of
magnetic field and dot-size, just as intuitively expected. It is not difficult to demonstrate analytically that
the Fermi energy becomes asymptotic to the x-axis -- and hence attains a zero slope -- at large values of $B$
and $R$. To this end, while the $B$-dependence in Eq. (10) is obvious, the $R$-dependence is not. To be
laconic, $\alpha_{nl}$, which is determined by enforcing the proper boundary condition [see Sec. II.A],
diminishes at large $B$ as well as at large $R$.

\begin{figure}[htbp]
\includegraphics*[width=8.0cm,height=9.0cm]{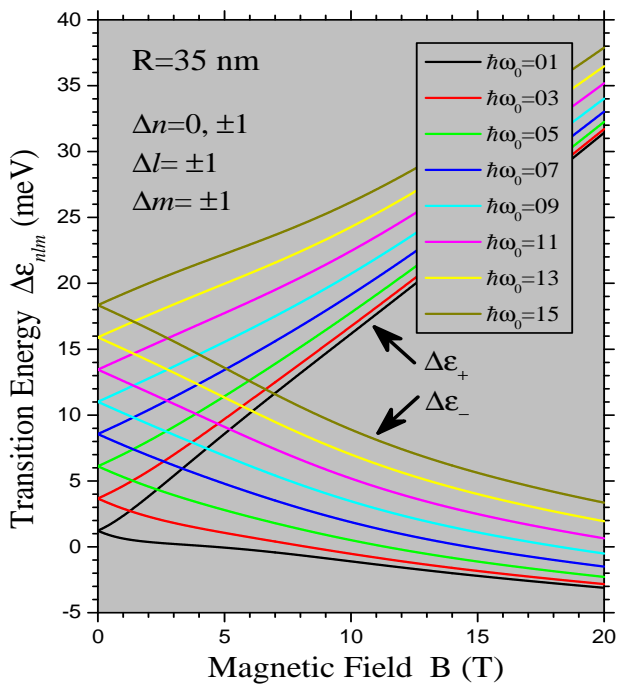}
\caption{(Color online) Allowed magneto-optical transitions as a function of the magnetic field ($B$), for
several values of the confinement potential $\hbar\omega_o$. The dot radius is defined as $R=35$ nm. The
selection rules are listed in the picture. The upper (lower) transition is designated as $\Delta\epsilon_+$
($\Delta\epsilon_-$).}
\label{fig12}
\end{figure}

Figure 12 describes the magneto-optical transitions in the spherical quantum dots as a function of magnetic
field for a given value of the dot size ($R=35$ nm), for several values of the confinement potential
($01\le \hbar\omega_o$ (meV) $\le 15$). Figure 12 is actually based on the computation of the exact
analytical expression for the transition energy in Eq. (39). The selection rules derived in Sec. II.C are as
listed inside the picture. Some simple mathematical manipulations of Eq. (39) reveal that, unlike the
laterally confined quantum dots [2, 51], both transitions -- upper [$\Delta\epsilon_+$] and
lower [$\Delta\epsilon_-$] -- will always survive whether or not $\hbar\omega_o=0$. Similarly, at $B=0$, we
are still left with two, albeit relatively weaker, transitions given
by $\Delta\epsilon_{\pm}=2 [\Delta\alpha \pm \frac{1}{2}]\,\hbar\omega_o$.
Even when $B\rightarrow\infty$ -- i.e., $\omega_c\gg\omega_o$ -- we cannot avoid either of the two
transitions. In other words, the SQDs do not allow the intra-Landau level transitions. The presence of the
parameters $\sigma$ and $\alpha_{nl}$ disallow the edges of the wedges to be exactly characterized by the
confinement potential ($\hbar\omega_o$) at $B=0$. These findings should encourage magneto-optical experiments
aimed at verifying such details in SQDs.

\begin{figure}[htbp]
\includegraphics*[width=8.0cm,height=9.0cm]{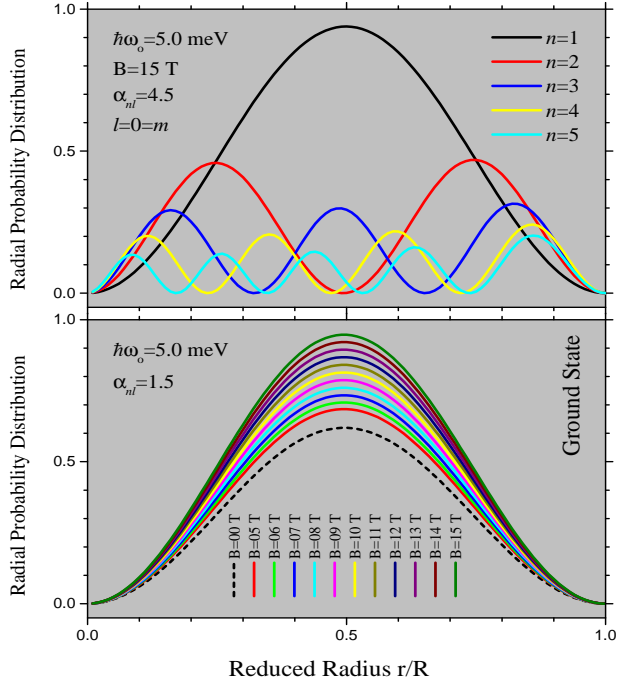}
\caption{(Color online) The radial probability distribution $Q_{nl}$($r$) in the ground state for the GaAs
SQDs against the reduced (radial) coordinate $r/R$. The confinement potential $\hbar\omega_o = 5$ meV). The
magnetic field $B=15$ T in the upper panel for the ground state, for several values of the principal quantum
number $n$, which distinguishes the $s$ orbitals characterized by $l=0$. The number of nodes equals ($n-1$).
The lower panel shows $Q_{nl}$($r$) in the ground state ($n=1,\, l=0,\, m=0$) for several values of magnetic
field $B$.}
\label{fig13}
\end{figure}

Figure 13 shows the radial probability distribution in the ground state for the GaAs SQDs against the reduced
radial coordinate $r/R$ for a given confinement potential ($\hbar\omega_o=5$ meV). The radial probability
distribution $Q_{nl}$($r$) defined
as $Q_{nl}(r)=\int^{\pi}_{0} d\theta\,\sin\theta\,\int^{2\pi}_0 d\phi\,r^2\,|\Psi(r, \theta, \phi)|^2$
expresses
the probability of finding an electron as a function of $r$ at a given instant. Before we proceed further, it
is worthwhile to remember that the (involved) associated Legendre function $P^m_l(x)=1$ for the quantum
numbers $l=0=m$, irrespective of its argument (x). The upper panel plots the ground states ($l=0,\,m=0$) along
with
$n=1$, $n=2$, $n=3$, $n=4$, and $n=5$, for the magnetic field $B=15$ T. In the language of {\em spectroscopy},
the principal quantum $n$ distinguishes the $s$ orbitals -- characterized by $l=0$. The number of nodes -- i.e.,
the values of $r$ where the $Q_{nl}$($r$) is zero -- equals ($n-1$).

The lower panel of Fig. 13 exhibits the probability distribution in the ground state for several values of the
magnetic field in the range $0\le B$\,(T)\,$\le15$. Note that $Q_{nl}$($r$) is zero at $r=0$ because the volume
of space available $r^2\,dr=0$. As $r$ increases, the dot size increases and so does $Q_{nl}$($r$). However,
the probability density $|\Psi(r, \theta, \phi)|^2$ decreases with increasing $r$, which implies that the $Q_{nl}$($r$)-path must have observed a maximum, where the probability of locating the electron is eminent. The
influence of an applied magnetic field on the variation of $Q_{nl}$($r$) is interesting to notice. We find that
the magnetic field tends to further add to the confinement and hence maximize the probability distribution with
increasing $B$. The crux of the matter is that the magnetic field does not shift the peak position of the radial distribution, which lies at $r/R\simeq 0.495$ for the whole range of $B$, including $B=0$.

\section{Concluding remarks}

In summary, we thoroughly investigated the magneto-optical absorption in the colloidally prepared spherical GaAs
quantum dots in the presence of a confining harmonic potential and an applied magnetic field in the symmetric
gauge. The theoretical formulation is constituted within the framework of Bohm-Pines' full-fledged RPA enabling
us to derive the Dyson equation for the total (or interacting) density-density correlation function. For this
purpose, we employ the exact single-particle eigenfunctions and eigenenergies obtained in Sec. II.A and the
strategy of determining the non-local, dynamic {\em inverse} dielectric function from Ref. 49. This served us to
analyze the Fermi's golden rule in order to derive the selection rules for the magneto-optical transitions and
extract the right expression for the radial probability distribution for the SQDs. Studying the size-dependence
of the Fermi energy for the SQDs was a part of this theoretical goal. All illustrative examples discussed here
are thus the outcomes of the computation based on the exact analytical results.

The magneto-optical absorption peak is blue-shifted with decreasing dot-size, increasing confinement, and with
increasing magnetic field. The larger dot-size and stronger magnetic field can enable us to debate the gradual
transitions from pure spatial to pure magnetic quantization. At lower dot-size and weaker magnetic field, the
Fermi energy decreases with increasing size and field. For large dot-size and strong magnetic fields, the Fermi
energy makes saw-tooth-like oscillations and finally becomes insensitive to both of them. It is not difficult
to justify this behavior analytically. The SQDs always allow both (upper and lower) transitions to survive even
in the extreme cases of vanishing confinement or magnetic field. While the magnetic field tends to maximize the probability of finding the electron, it does not shift the peak position of the radial distribution. These
findings should motivate the magneto-optical experiments on SQDs aimed at verifying such details as listed here.

What is so special about the spherical geometry (SG)? It is a historical fact that the SG has played a model role
for testing various classical and quantal hypotheses in the past. One of the latest instances is the use of the
SG for studying the fractional quantum Hall effect (FQHE) [52]. Here a 2D sheet of electrons is wrapped up on a
spherical surface exposed to a radial magnetic field generated by a Dirac magnetic monopole placed at the center
of the sphere. The motivations behind the SG are two-fold [53]: First, it is free from edges, which makes it
potentially fit for studying the bulk properties. Second, Landau levels (LLs) have a finite degeneracy (for a
finite magnetic field), which helps discern incompressible states for finite systems. The SG has been pivotal in constructing the theory of FQHE and in furnishing proofs for numerous related features.

Despite the in-depth research pursuits of electronic, optical, and transport phenomena in quantum dots in the
past, our practical knowledge required to harness the full potential of these man-made atoms still seems to be
in its infancy. It has been instructive to inquire into the addition spectra using the concept of
{\em Coulomb blockade} that helps provide control over the number of electrons by adjusting the energy required.
The exchange interaction plays a role and the energy needed to add 
a second electron in the presence of an existing one also depends upon their relative spin.
The long life-time of the (restricted) electron-electron interactions commonly gives rise to
{\em decoherence}, 
implying that the dots can offer a potential system for exploring technologies based upon the quantum coherence.
As an example, it is likely to create and control superimposed or even {\em entangled} states using the highly
coherent lasers. External control over the full quantum eigenfunction in a semiconducting SQD may even lead to
exciting applications such as those involving quantum computing.


\begin{acknowledgments}
The author would like to express his sincere thanks to Loren Pfeiffer, Daniel Gammon, and Aron Pinczuk
for the very useful communication regarding the fabrication of the spherical quantum dots. He would
also like to thank Peter Nordlander, Naomi Halas, and Thomas Killian for all the support and
encouragement. He appreciates Kevin Singh for the unconditional help with the software.
\end{acknowledgments}

\newpage

\end{document}